\documentclass{article}

\begin{document}

\title{Universality classes for self-similarity  of noiseless
multi-dimensional Burgers turbulence and interface growth}
\author{S.N. Gurbatov\\
 Radiophysics Department, University of Nizhny Novgorod,\\
23 Gagarin Ave., 603600, Russia (permanent address)\\
CNRS UMR 6529, Observatoire de la C\^ote d'Azur, Lab.G.D.Cassini\\ 
B.P. 4229, 06304  Nice Cedex 4, France}

\date{}

\maketitle

\begin{abstract} 
The present work is devoted to the evolution of random solutions of
the unforced Burgers and KPZ equations in d--dimensions in the limit 
of vanishing viscosity. We consider a cellular model and as 
initial condition  assign a value for the velocity potential chosen 
independently within each cell. We
show that the asymptotic behavior of the turbulence at large times 
is determined by the tail of the initial potential 
probability distribution function.
Three classes of initial distribution
leading to self--similar evolution are identified:
(a) distributions with a power--law tail,  
(b) compactly supported potential,  
(c) stretched exponential tails.
In class (c) we find that the mean potential
(mean height of the surface) 
increases logarithmically with time and the ``turbulence energy'' 
$E(t)=\langle {\bf v}^2({\bf x},t)\rangle=
\langle (\nabla {\psi}({\bf x},t))^2\rangle$ 
(mean square gradient of the surface)  decays as
$ t^{-1}$ times a logarithmic correction.
In classes (a) and (b) we find that the changes in the  mean
potential   and energy  have a power-law time dependence, namely
$E(t)\propto t^{-p}$ where the index $p$
lies in the interval $2>p>(2-d)/2$.  
In class (c) the roughness of the surface, measured 
by its mean--square gradient, may either decrease  or increase with time.
We discuss also the influence of finite viscosity and long range correlation 
on the late stage evolution of the Burgers turbulence.
\end{abstract}
   
PACS number(s)\,: 47.27.Gs, 
                        02.50.Ey 

\newpage

\section{Introduction}

The multi-dimensional Burgers equation
\begin{equation}
\label{Burgd}
\frac{\partial {\bf v}}{\partial t} + \left(
{\bf v}\cdot\nabla {\bf v} \right) = \nu \triangle {\bf v}\,,
\end{equation}
is a generalization of the well--known Burgers equation
\begin{equation}
\label{Burg1}
\frac{\partial v}{\partial t} + v\,\frac{\partial v}{\partial x} = \nu
\frac{\partial^2 v}{\partial x^2}\,.
\end{equation}

The  nonlinear diffusion equation (\ref{Burg1}), which 
was originally introduced by J.M.Burgers in 1939  as 
a model of hydrodynamical turbulence \cite{Burgers1939,Burgers1974}, 
was later shown  to describe a
variety of nonlinear wave phenomena arising in the theory of wave
propagation, acoustics, plasma physics (see, e.g.,
\cite{Whitham,RudenkoSoluyan,GMS91}).

The Burgers equation (\ref{Burg1}) describes two principal
effects inherent in  any turbulence \cite{Frisch}: the nonlinear
redistribution of energy over the spectrum and the
action of viscosity in  small-scale regions.
Although  external forces
are not present in (\ref{Burg1}), the one-dimensional Burgers equation
does describe the decay of turbulence, i.e. the nonlinear
transformation of the random initial field ${\ v}_0(x)$.
The Burgers equation shares a number of 
properties with the Navier--Stokes
equation, namely the same type of nonlinearity, of invariance groups, of the
energy-dissipation relation, of the existence of a multidimensional version,
etc.  However, Burgers equation  is integrable and 
therefore is not sensitively dependent 
on initial conditions. 
The differences between the Burgers  and  Navier-Stokes equations
are as interesting as the similarities \cite{Kr68} and this is also
true for the multi-dimensional Burgers equation (\ref{Burgd}).

The three--dimensional form of  equation (\ref{Burgd}) has 
been used to model the  formation of the large scale 
structure of the Universe when  pressure is 
negligible, that is, during the nonlinear stage of the gravitational 
instability arising from random initial perturbations 
\cite{GurbatovSaichev1984,GSS89,SZ89,VDFN94}.
Other problems leading to the multi--dimensional Burgers equation, 
or variants of it, include surface growth under sputter deposition and 
flame front motion 
\cite{BS95}. In such instances, the potential ${\psi}$
corresponds to the shape of the front's surface, and
the equation for the velocity potential ${\psi}$ is 
identical to the KPZ (Kardar, Parisi, Zhang) equation
\cite{BS95,KardarParisiZhang,WW98}. For the deposition problem
${\bf v}=-\nabla\psi$ is the gradient of the surface. The roughness 
of the surface, measured by its mean--square gradient
$E(t)=\langle(\nabla {\psi}({\bf x},t))^2\rangle=
\langle {\bf v}^2({\bf x},t)\rangle$,  may either decrease 
or increase with time. Nevertheless we will use the expression 
``turbulence energy'' for this value of $E(t)$. 
With  external random forces the
Burgers and KPZ equations describe  
phenomena such  as turbulence without pressure, disordered systems, 
directed polymers, etc 
\cite{BS95,BMP95,CheklovYakhot,Polyakov,EKMS}.

Here we will consider the evolution of the  velocity field 
${\bf v}({\bf r},t)$ and potential ${{\psi}({\bf r},t)}$ as given by the
noiseless multi--dimensional Burgers and KPZ equations.
In this case the evolution of the fields is
fully determined by the statistical properties of the initial field
${\bf v}_0({\bf r})$, which is assumed to be potential 
${\bf v}_0({\bf r})=-\nabla\psi_0({\bf r})$.
Our main attention will be given to  the
case of vanishing viscosity $(\nu \rightarrow 0)$, when the dissipative
effects are important only in the vicinity of  shocks.

It is known that the asymptotic behavior $(t\rightarrow\infty)$ of 
Burgers turbulence  strongly depends on the behavior of the structure
function of the initial potential $d_\psi ({\bf \rho})=\langle(\psi_0
({\bf r}+{\bf \rho})-\psi({\bf r}))^2
\rangle$ at large distances \cite{Burgers1974,GMS91}.
If the structure function $d_\psi ({\bf \rho})$ increases as a power
law in space then the initial potential field is Brownian, or 
fractional Brownian  motion, and some scaling may be used
\cite{Burgers1974,GMS91,VDFN94,SheAurellFrisch}.
In this case the turbulence is self-similar. 
The evolution of the external scale $L(t)$ 
of the self--similar solution in time is
determined by the spatial behavior of $d_\psi ({\bf \rho})$ 
at large distances. Very recently a 
complete solution of the one--dimensional Burgers
equation, with initial Gaussian white noise distributed data in the 
inviscid limit, has been  obtained \cite{FrMart99}.

If the structure function of the potential is bounded at
${\bf \rho}\rightarrow \infty$, then  scaling 
arguments no longer work. Kida  \cite{Kida} has proposed  using, 
for the one-dimensional Burgers  
turbulence, a discrete cellular model of the initial conditions with
independent distribution of potential in different cells. 
He has shown that  the energy decays like  $t^{-1}$
with a logarithmic correction when the probability distribution of the
potential in each cell has a stretched 
exponential tail $\sim \exp(-H^\beta)$.
Later it was shown \cite{GurbatovSaichev1981,FournierFrisch,MSW95} 
that for an initial continuous Gaussian  field all of the
statistical characteristics of  one--dimensional Burgers  turbulence
become   self--similar and the energy decays as
$t^{-1}/(\ln t)^{\frac{1}{2}}$.
Several models of the evolution of Burgers turbulence
with an initial perturbation of non Gaussian type have   
also been proposed (\cite{WW98} and references therein, 
\cite{AMS94,EsipovNewman,Esipov,Newman97,BeKr98}). It has been shown 
that the law of energy decay 
strongly depends on the statistical properties of the initial field  
with  homogeneous potential.

It is known that  in the limit of vanishing viscosity 
the solution of Burgers' equation is reduced to 
searching for the absolute maximum of some function of the initial
potential \cite{Hopf}. Consequently, the statistical properties  
of  Burgers'  turbulence are
determined by the behavior of the probability distribution of the initial
potential.
One of the important results of the
classical theory of  extrema is that there
are only three universal classes of ``extreme value distributions'' 
of the  sequence of
independent and identically distributed random
variables \cite{LeadbetterLindgrenRootzen}. This result was used in 
\cite{BM97} for dealing with the    problem of the 
equilibrium of low-temperature physics of disordered systems and partly for 
the one-dimensional decaying Burgers turbulence.
One of our main tasks  is to show that, in the 
discrete cellular model of the initial condition for the d-dimensional
Burgers turbulence, there are also three classes of universal self-similar
evolution of the velocity and potential fields. 

The paper is organized as follows. In Section 2  we
formulate our problem and list some elementary results about the Burgers
equation. We also consider  the evolution of 
potential and velocity fields for a simple kind of perturbation 
in individual cells at the initial stage. 
In section 3 we derive  the general 
expression for the energy and probability distribution 
functions of the potential and vector velocity fields. In
Section 4 we show that three types of  initial condition 
lead to asymptotic self-similar behavior of Burgers' turbulence. 
Section 5 presents concluding remarks. We also discuss  the
influence of finite viscosity and long range correlation on the late 
stage evolution of Burgers' turbulence.

\section{Description of the basic model}
\subsection{Basic equations and  local self--similarity}
\label{s:basic}

We will discuss the initial-value problem for the unforced Burgers equation
(\ref{Burgd}), and consider only the potential solution of this equation,
namely
\begin{equation}
{\bf v}({\bf x},t) =-\nabla\psi({\bf x},t)\,.
\label{vpsi}
\end{equation}
The velocity potential $\psi({\bf x},t)$ satisfies 
the following nonlinear equation:
\begin{equation} 
\partial_t{\psi}=\frac{1}{2}(\nabla\psi)^2+\nu\triangle\psi \,.
\label{psieq}
\end{equation}

This equation is the same as the KPZ
equation \cite{BS95,KardarParisiZhang,WW98}, which is usually 
written in  terms of the variable $h={\lambda}^{-1}\psi$. The parameter 
$\lambda$ has the dimensions of length divided by time and is the local 
velocity of the surface growth. Henceforth $h$ has the 
dimensionality of length and is the measure of shape of the surface.
Using the Hopf--Cole transformation $\psi=\ln{U}$ \cite{Hopf,Cole},
one can reduce (\ref{psieq})  to a  linear diffusion equation.
We are mainly interested here by solutions in the limit 
$\nu\rightarrow{0}$. Use of Laplace's method then
leads to the following ``maximum representation''  for the potential and
velocity fields in the limit of vanishing viscosity
\cite{Hopf,GMS91,VDFN94}:                         
\begin{eqnarray}
\psi({\bf x},t)=\max_{{\bf y}}\Phi({\bf x},{\bf y},t)\,,  \\
\Phi({\bf x},{\bf y},t)=
\psi_0({\bf y})-\frac{({\bf x}-{\bf y})^2}{2t}\,,
\label{psimax}
\end{eqnarray}   
\begin{equation}
{\bf v }({\bf x},t)=\frac{{\bf x}-{\bf y}({\bf x},t)}{t}\,.
\label{vxy}
\end{equation}
Here $\psi_0({\bf y})$ is the initial potential  and 
${\bf v}_0({\bf x})=-\nabla\psi_0({\bf x})$.
In (\ref{vxy}) ${\bf y}({\bf x},t)$ is the Lagrangian coordinate 
where the function 
$\Phi({\bf x},{\bf y},t)$ achieves its global or absolute maximum 
for a given coordinate ${\bf x}$ and time $t$.

At large times the paraboloid peak in (\ref{psimax}) defines a 
much smoother function than the initial potential $\psi_0({\bf y})$.
Consequently, the absolute maximum of $\Phi({\bf x},{\bf y},t)$
coincides with one of the local maxima of $\psi_0({\bf y})$.  
The Lagrangian coordinate  ${\bf y}({\bf x},t)$ then
becomes  a discontinuous function of ${\bf x}$, constant within a cell, 
but jumping at the boundaries \cite{GMS91,VDFN94}. In each
cell fluid particles move away from the small region near the cell
center ${\bf y}_k$. The velocity field ${\bf v}({\bf x},t)$ has 
discontinuities (shocks) and the potential field $\psi({\bf x},t)$
has   gradient discontinuities (cusps) at the cell boundaries;
these shock surfaces or walls form a connected structure.
Inside the cells the velocity and potential fields (surface shape) have a
universal self-similar structure:
\begin{equation}
{\bf v} ({\bf x},t)=\frac{{\bf x}-{\bf y}_k}{t}\,,
\label{vxy_k}
\end{equation}
\begin{equation}
\psi({\bf x},t)=\psi_0({\bf y}_k)-\frac{({\bf x}-{\bf y}_k)^2}{2t}\,.
\label{psimax_k}
\end{equation}
The longitudinal component of the velocity vector ${\bf v} ({\bf x},t)$ 
consists of a sequence of sawtooth pulses with 
random positions of the shocks 
and ``zeros'', just as in one dimension. The transverse components 
consist of sequences of rectangular pulses with random amplitudes and 
random positions of the shocks. Wall motion results in continuous 
change of cell shape with cells swallowing their neighbors and thereby 
inducing growth of the external scale $L(t)$ of the Burgers  turbulence.
At large times the behavior of the turbulence will be  
determined by the statistical properties of the initial potential 
and, moreover,
by the statistical properties of local maxima $\psi_0({\bf y}_k)$. 

\subsection{The cellular model and initial stage of evolution  }
\label{s:initial}
                
We consider the cellular model of the initial conditions, in
which we assume that  space is divided  into identical  
cells, each having  a volume $dV=L_0^d$, where $L_0$ is
the length of the cell and $d$ is the spatial dimension. We assign
an initial value for the potential which is chosen 
independently within each cell.
The same approach was used  in \cite{Kida} in order to find the 
probability distributions of the  amplitudes and velocities of shocks
for the  one--dimensional Burgers turbulence and 
in \cite{GMS91,GurbatovSaichev1984} 
for  multi-dimensional Burgers turbulence 
in the case where the initial potential had a
a stretched exponential tail. 
In the ``shot--noise model'' \cite {AMS94}
it is assumed that the initial potential is a sum of 
potentials of ``unit non-homogeneities''  with random amplitudes and
scales and with a Poisson ensemble for the  
position.

First we consider the evolution of the velocity and potential 
fields of the  ``unit non-homogeneities'' inside the cells assuming that 
there is no interaction between  cells.
We  assume for  simplicity that inside each cell the initial 
potential and velocity are isotropic and
\begin{equation}
\psi_0({\bf x})=\psi_0(1-{\bf x} ^2/l_0^2)\,,\;\;\;|{\bf x}|<l_0\,,
\label{Psi0}
\end{equation}
\begin{equation}
{\bf v}_0({\bf x})=V_0 {\bf x}/l_0\,,\;\;\;|{\bf x}|<l_0
\,,\;V_0=2\psi_0/l_0\,,
\label{v0}
\end{equation}
where $\psi_0({\bf x})=0,{\bf v}_0({\bf x})=0$ when $|{\bf x}|>l_0$ and
$V_0$ is the amplitude of the velocity at the border $|{\bf x}|=l_0$.
From the solution (\ref{psimax},\ref{vxy})  we have for the velocity field
\begin{equation}
{\bf v}({\bf x},t)= 
V_0\displaystyle\frac{{\bf x}}{l_0(1+V_0t/l_0)}\,,\;\;
|{\bf x}|<x_s\,,
\label{vxt}
\end{equation}
where $x_s$ is the position of the shock surface
\begin{equation}
x_s=l_0(1+V_0t/l_0)^{1/2}=(l_0^2+2\psi_0 t)^{1/2}\,.
\label{xshock}
\end{equation}
In the one-dimensional case this solution is a 
well known N-wave \cite{Whitham}.

The energy of the velocity field is given by
\begin{equation}
E(t)=\int v^2({\bf x},t)d^d{\bf x}=\\
\displaystyle\frac{2\pi^{d/2}}{\Gamma(d/2)}
\int\limits_0^{x_s} v^2  r^{d-1}dr\,,
\label{Edef}
\end{equation}
and from (\ref{vxt},\ref{xshock}) it follows that
\begin{equation}
E(t)=E_0\left(1+V_0 t/l_0\right)^{\frac{d-2}{2}}\,,\;\;
E_0=\displaystyle\frac{2\pi^{d/2}V_0^2 l_0^d}{\Gamma(d/2)(d+2)}\,,
\label{Et}
\end{equation}
where $\Gamma(z)$ is a gamma--function.

From (\ref{Et}) we see that for $d=1$ 
the energy $E(t)$ decreases with time, for $d=2$ $E(t)$ is constant, 
and for $d\ge 3$ $E(t)$  increases with time.
It is possible to show that $E(t)$ may increase with time when
$d\ge 3$ even if the  viscosity coefficient $\nu$ is finite.
The increase of the energy in the  multi-dimensional Burgers equation (in
contrast with $d=1$) is the result of  this equation  not having a
conservation form. 

Moreover it easy to show that in the multidimensional Burgers equation
the energy does not conserve in the limit of vanishing viscosity
even at the initial stage of
evolution, when the velocity field does not have shocks.  Let us consider
the evolution of the continuos velocity field having only a radial
component $v_r ({\bf x},t)=v(r,t)$.  In the limit of vanishing
viscosity we have the implicit solution of Burgers
equation $v=v_0(r- vt)$.  Substitute this solution
into (\ref{Edef}) and replace
the Eulerian coordinate $r$ by Lagrangian coordinate $y$:
$r=y+tv_0 (y)$, we have the following expression for the energy
\begin{equation}
E(t)=\displaystyle\frac{2\pi^{d/2}}{\Gamma(d/2)}
\int\limits_0^{\infty} v_0^2 (y) (y+tv_0 (y))^{d-1}
\left(1+t\frac{\partial v_0(y)}{\partial y}\right)dy\,.
\label{Edefr}
\end{equation}
From (\ref{Edefr}) we see that the energy $E(t)$ of continuos
field conserves only in one--dimensional case ($d=1$). In particular
for $d=2$ we have from (\ref{Edefr})
\begin{equation}
E(t)=2\pi\int\limits_0^{\infty} v_0^2(y)dy+ t
\displaystyle\frac{4\pi}{3} \int\limits_0^{\infty}  v_0^3(y)dy\,.
\label{Edef2}
\end{equation}
Thus for the positive perturbation of the initial velocity
the energy (mean--square gradient of the surface) increases
with time, and for the negative perturbation of the velocity
the energy decreases with time. This effect takes place due to the
relatively more significant contribution into mean--square gradient
of the surface the regions with large distances $r$ (\ref{Edefr}).

The asymptotic behavior of the potential and velocity fields is determined
only by the positions and values of the maximum  of the initial 
potential in the cells (\ref{vxy_k},\ref{psimax_k}).
For  relatively large time $(V_0 t\gg l_0)$  the velocity
inside the cells has the universal behavior
\begin{equation}
{\bf v}({\bf x},t)={\bf x} /t\,,\;\;|{\bf x}|\le x_s=(2\psi_0 t)^{1/2}\,.
\label{vxtas}
\end{equation}
The position of the shock surface $x_s$  
and the energy of the field
\begin{equation}
E(t)\sim\psi_0^{\frac{d+2}{2}}t^{\frac{d-2}{2}}
\label{Etas}
\end{equation}
are determined only by the value of the initial potential maximum $\psi_0$.
It is easy to show that this asymptotic behavior takes place 
for the arbitrary initial localized
perturbations with the potential maximum $\psi_0$.

Let us assume ($L_0\gg l_0$) and that the
maximum value in the  ``$i$-th'' cell is $\psi_{0,i}$.
The interaction between  cells begins when the 
shock surfaces of neighboring
cells come in contact, i.e. when $|x_{s,i}|+|x_{s,j}|=L_0$.
At rather large time the border between the cells becomes  a hyper--plane, 
moving with constant velocity 
$|\psi_{0,i}-\psi_{0,j}|/|{\bf y}_i-{\bf y}_j|$ away
from the cell with a larger value of the unital potential \cite{GMS91}.
When the potential maximum in all of the cells is the same, 
i.e. when  $\psi_{0,i}=\psi_0$,
then at large times the hyper--surfaces between neighboring cells  
are immobile and the velocity ${\bf v}({\bf x},t)$ and potential 
${\psi}({\bf x},t)$ fields have stationary periodic structure. 
The gradient of the velocity field decreases as $t^{-1}$, 
which means that  the energy density $E(t)=\langle {\bf v}^2\rangle$ 
decreases according to
\begin{equation}
E(t)\sim L^2_0 /t^2\,.
\label{Etper}
\end{equation}

For a random distribution of the potential maximum $\psi_{0,i}$ we have
the permanent absorption of cells with low values of the potential so that
the external scale of the turbulence $L(t)$ increases with time.
The mean energy evolution strongly depends on the statistical properties of
the maximum potential distribution. The energy decay  is slower 
than for the periodic structure (\ref{Etper}). The interaction between
cells changes also the law (\ref{Etas}). Moreover, for $d \ge 3$, 
due to the interaction between  cells   the
increase of energy $E(t)$ (\ref{Etas}) may change to a to a decrease.
It will be shown later that asymptotically 
$E(t)\propto t^{-p}$ with $p$
lyings in the interval $2>p>(2-d)/2$.

\section{Energy  evolution  and probability distributions 
of the turbulence and the interface}
\label{s:energy}

We assume that the initial random potential $\psi_0({\bf y})$ is
statistically homogeneous. Then, from equation (\ref{psieq}) it 
follows that the turbulence energy  is determined by 
the time derivative of the mean potential
\begin{equation}
E(t)=\langle {\bf v}^2({\bf x},t)\rangle=\langle(\nabla 
{\psi}({\bf x},t))^2\rangle=
2\,\frac{\partial\langle\psi({\bf x},t)\rangle}{\partial t}\,.
\label{ephi}
\end{equation}
Let  $P_{\psi}(H,t)$ denote the  probability density of $\psi$ at time
$t$ and let  $Q_{\psi}(H,t)$ denote the cumulative probability to 
have $\psi<H$, given by
\begin{equation}
Q_{\psi}(H,t)=\int\limits_{-\infty}^H P_{\psi}(A,t)dA\,.
\label{qp}
\end{equation}
The mean value of the potential $\langle\psi\rangle$ 
(i.e. mean value of the height of the surface) 
at time $t$ is then expressible as
\begin{equation}
\langle\psi\rangle=\int\limits_{-\infty}^{+\infty} HP_{\psi}(H,t)dH\,.
\label{meanpsi}
\end{equation}
Thus we can find the mean energy  of the turbulence  if we know the
probability distribution function $P_{\psi}(H,t)$, 
where $P_{\psi}(H,t)dH$
is the probability that the absolute maximum of $\Phi({\bf x},{\bf y},t)$
(\ref{psimax}) lies in the interval $H,H+dH$ for all ${\bf y}$.

The function  $P_{\psi,{\bf y}}(H,{\bf y};{\bf x},t)$   
is the joint probability distribution function (p.d.f.)
of $\psi$ and ${\bf y}$ and  $P_{\psi}dHd^d{\bf y}$ is the
the probability that the absolute maximum of 
$\Phi({\bf x},{\bf y},t)$ lies in the interval 
$H,H+dH$ in the volume element $d^d {\bf y}$.  
By integrating the
$P_{\psi,{\bf y}}(H,{\bf y};{\bf x},t)$ over $H$ we obtain the p.d.f.
of the Lagrangian coordinate {\bf y}:
\begin{equation}
P_{{\bf y}}({\bf y};{\bf x},t)=\int\limits_{-\infty}^{+\infty}
P_{\psi,{\bf y}}(H,{\bf y};{\bf x},t)dH\,,
\label{py}
\end{equation}
which   permits one to find the probability distribution
function of the velocity field (\ref{vxy}).

Since the initial potential is homogeneous we will consider the
point ${\bf x}=0$ and no longer include  ${\bf x}$ in the 
parameter of the probability distribution function.
In the ``cellular''  model we assume that the space is
divided up into elementary cells of  volume $dV=L_0^d$. 
We consider the late stage of evolution when the
external scale $L(t)$ is much greater then 
$L_0$. On this scale the initial potential field is homogeneous.
(Formally we can introduce a random uniform distribution of cell positions.)

The asymptotic behavior of the probability distribution
function of absolute maximum of a large number of random quantities is
determined by the asymptotic properties of their cumulative distribution
function at large values. We assume that in each cell the cumulative
probability function of the initial potential $\psi_0({\bf y})$ is 
the same and can be represented in the following form:
\begin{equation}
Prob(\psi_0<H)=F(H)=1-f(H)\,.
\label{fmax}
\end{equation}
The cumulative probability distribution function for  $\Phi)$
(\ref{psimax}) in the $i$-th cell is
\begin{equation}
F_i(H)=1-f_i(H) \,,\;\;
f_i(H)=f(H+\frac{{\bf y}_i^2}{2t})\,.
\label{fertmax}
\end{equation}
Here ${\bf y}_i$ is the coordinate of the $i$-th cell.
Due to the independence of the initial potential in different cells we can
determine the cumulative distribution function $\psi$ as
\begin{eqnarray}
Prob(\psi<H)=Q(H,t)=\prod F_i(H)
\nonumber\\\
\equiv\prod(1-f_i(H))\,.
\label{qm}
\end{eqnarray}
From (\ref{fertmax}) we see that $f_i(H)$ decreases with the distance
$|{\bf y}|$.

We will consider the limit when $t$ becomes large, 
so that the  number of factors 
in (\ref{qm}) becomes significant.
This means that the absolute maximum of $\Phi$ is reached for
rather large $H$, so in each cell we have $f_i(H)<<1$, 
thus $1-f_i(H)\simeq\exp{[-f_i(H)]}$, 
and we can rewrite (\ref{qm}) in the form
\begin{equation}
Q(H,t)=e^{-N(H,t)}\,,
\label{qpp}
\end{equation}
where
\begin{equation}
N(H,t)=\sum_i f_i(H)=\sum_i f(H+\frac{{\bf y}_i^2}{2t})\,.
\label{Nsum}
\end{equation}
At large $t$ when the difference between $f_i$ in neighboring cells is small,
we can replace the summation in (\ref{Nsum}) by an 
a volume integral:
\begin{equation}
N(H,t)=\frac{1}{L_0^d}\int f(H+\frac{{\bf y}^2}{2t})d^d {\bf y}\,.
\label{nint}
\end{equation}
For large $H$ the events for which $\psi_i$ is greater than $H$ are rare
and they have a Poisson distribution.
Thus N in (\ref{qpp}) is a mean number of events when $\psi>H$.

From (\ref{qpp},\ref{nint}) we have two equivalent expressions for the
mean potential:
\begin{equation}
\langle\psi\rangle=
\int\limits_{-\infty}^{+\infty}H \frac{\partial}{\partial H}
e^{-N(H,t)}dH=\int\limits_{0}^{+\infty}H(\Theta,t) e^{-\Theta}d\Theta\,,
\label{psimean1}
\end{equation}
where $H(\Theta,t)$ is a solution of the equation
\begin{equation}
N(H,t)=\Theta\,.
\label{nteta}
\end{equation}
Formally we can integrate in (\ref{psimean1})  over 
the infinite interval of $H$ since the
probability of  $H$ being  negative is negligibly small  
when   $t$ is large.

In order to find the probability distribution function of 
the velocity we first of all need to 
find  the joint probability distribution of value and the
coordinate of the absolute maximum of $\Phi$.
The probability that the absolute maximum is in the $i$-th cell, 
with a value in the interval $H,H+dH$,  is
\begin{eqnarray}
Prob({\bf y} \in \triangle_j, H\in[H,H+dH])
\nonumber\\=-\frac{\partial}{\partial H}
f_j(H)\prod_{i \ne j}(1-f_i(H))\,.
\end{eqnarray}
Here the product over all cells with $i \ne j$ is the probability that
outside the $j$-th cell all local maxima are less than $H$.

With the same assumption as before we can find the joint probability
distribution coordinate and value of the absolute maximum.
At large $t$ we have 
\begin{equation}
P_{y}({\bf y},t)=P({\bf y} \in \triangle_j)/L_0^d\,.
\end{equation}
and after  integration over $H$ we have the following expression for
the probability distribution function of the absolute maximum coordinate
\begin{equation}
P_{y}({\bf y},t)=-\frac{1}{L_0^d}\int\limits_{-\infty}^{+\infty}
\frac{\partial}{\partial H}f\left(H+\frac{{\bf y}^2}{2t}\right)
e^{-N(H,t)}dH\,.
\label{pyy}
\end{equation}
Integration by parts leads to
\begin{eqnarray}
P_{y}({\bf y},t)=\frac{1}{L_0^d}\int\limits_{-\infty}^{+\infty}
f\left(H+\frac{{\bf y}^2}{2t}\right)\frac{\partial}{\partial H}
e^{-N(H,t)}dH
\nonumber\\
=\frac{1}{L_0^d}\int\limits_{0}^{+\infty}
f\left(H(\Theta)+\frac{{\bf y}^2}{2t}\right)
e^{-\Theta}d\Theta\,.
\label{py1}
\end{eqnarray}
Here $H(\Theta)$ is a solution of equation (\ref{nteta}). 
Using (\ref{nint}) it is easy to see that the norm of $P_y({\bf y},t)$ 
is unity.

In the theory of the absolute maximum of random sequences 
\cite{LeadbetterLindgrenRootzen} the large parameter
of the theory is $M$ -- the number of  ``points'' in a sequence.
In our case we consider formally an  infinite number of  points, but owing
to the parabolic form used in (\ref{Nsum}) the effective number of  cells
is finite and increases with time. 
We stress again that 
expressions (\ref{nint},\ref{py1}) are valid only  when  $L(t)>>L_0$.

\section{Three types of self-similarity for  Burgers Turbulence}
\label{s:three}

In a previous section we derived  expressions for the probability
distribution function of the potential $\psi({\bf x},t)$
(\ref{qpp},\ref{nint}) and for the mean potential (\ref{psimean1}) 
which determines the
energy evolution  of the turbulence (\ref{ephi}). 
We also derived the form  
of  the probability
distribution function for the absolute maximum coordinate (\ref{pyy})
which  coincides with the form of the velocity probability distribution
function (\ref{vxy}). It is  known that there are three types, known as 
the Frechet, 
Weibull and Cumbel classes, of
asymptotic behavior for the probability distribution function of the large
sequence of  random quantities \cite{LeadbetterLindgrenRootzen,BM97}.
In this section we will show that these three types of initial condition
lead to  three types of asymptotic self-similar behavior of the Burgers
turbulence.

\subsection{Distributions with a power-law tail: Frechet class}
\label{s:frechet}
Let us now assume that the cumulative probability distribution function of
the initial potential (\ref{fmax}) has a power law tail
\begin{equation}
f(H)=\left( \frac{H_*}{H} \right)^\gamma \,.
\label{fhpow}
\end{equation}
In the  integral (\ref{nint}) for  $N(H,t)$  
we can integrate over the radius $r=|{\bf y}|$ and  obtain
\begin{equation}
N(H,t)=\frac{2\pi^{d/2}}{\Gamma(d/2)} \left( \frac{H_*}{H} \right)^\gamma
\int\limits_0^\infty \frac{r^{d-1}dr}{\left(1+\displaystyle
\frac{r^2}{2Ht}\right)^\gamma}\,.
\label{ng1}
\end{equation}

This integral converges when $\gamma>\frac d2$, and gives
the asymptotic distribution of the potential $\psi$ when the 
probability distribution function of the initial
potential decays sufficiently rapidly.
Replacing the variable $r$ by $r=(2tH)^{1/2}x$ allows us to write 
$N(H,t)$ as a power-law function of $H$
\begin{eqnarray}
N(H,t)=\frac{H_*^\gamma}{L_0^d}\frac{t^{d/2}}{H^{\frac{2\gamma-d}{2}}}C_{d} 
\,,
\nonumber\\
C_{d}=\frac{2^{\frac{d+1}{2}}\pi^{d/2}}{\Gamma(d/2)}\int\limits_0
^\infty \frac{x^{d-1}}{(1+x^2)^\gamma}dx\,.
\label{ngsel}
\end{eqnarray}
where $C_{d}$ depends on  $d$ as well as $\gamma$.
Thus we have the self-similar cumulative function (\ref{qpp}) 
of the potential (height of the surface) is given by
\begin{equation}
Q(H)=\exp\left[ -\left(\frac{H_*}{H}\right)^\frac{2\gamma-d}{2}
\left(\frac{t}{t_{nl}}\right)^{d/2}\right]\,.
\label{qrself}
\end{equation}
Here we introduce $t_{nl}$ as a nonlinear time of evolution
\begin{equation}
t_{nl}=\frac{L_0^2}{H_*C_{d}^{2/d}}\,.
\label{tnlr}
\end{equation}
It follows from  (\ref{qrself})  that the
probability distribution function of
potential is self-similar with the amplitude scale $H_s(t)$:
\begin{equation}
H_s(t)=H_*\left(\frac{t}{t_{nl}}\right)^{d/(2\gamma-d)} \,.
\label{hgself}
\end{equation}
The time evolution  of the velocity variance (\ref{ephi}) is determined by
the time derivative of the mean potential (\ref{psimean1}). It is easy
to see from (\ref{qrself}) that the expression for the  mean potential
converges if $\gamma>\displaystyle\frac d2+1$, and in this case the 
mean value $\langle \psi \rangle\sim H_s(t)$. 
For $\gamma < d$ the mean potential $\langle \psi \rangle$
increases faster in time then the local
height of the surface which grows proportionally $t\lambda$. The
superlinear increase of mean potential is possible while the KPZ
equation (\ref{psieq}) describes  the evolution of shape
${\psi}({\bf x},t))\sim h({\bf x},t)=H({\bf x},t)-t\lambda$ of the
surface and the normal growing of the surface is excluded a for this
equation \cite{BS95,KardarParisiZhang,WW98}.

From (\ref{ephi})  the velocity variance (turbulence energy)
is given by  
\begin{equation}
E(t)=2\frac{\partial \langle \psi\rangle}{\partial t} \sim \frac{\partial H_s}
{\partial t} \propto t^{-p} \,\,;\,\,\,\, p=\frac{2(\gamma-d)}{2\gamma-d}\,.
\label{eg}
\end{equation}
We stress here that for this expression to be valid $\gamma$ 
must satisfy   $\gamma > \gamma_{cr}
=\displaystyle\frac d2 +1$.
From (\ref{eg}) we see that when  $\gamma$ lies in the interval
\begin{equation}
\displaystyle\frac d2 +1 < \gamma < d\,,
\label{gdinc}
\end{equation}
the velocity variance  increases with time.
Thus the velocity variance increases only if the spatial dimension 
$d$ satisfies $d\geq 3$. For $d=3$  we have the same results as
the ``shot-noise model'' of the initial perturbation 
\cite{AMS94}.
On the left hand side  of the inequality (\ref{gdinc}), if $\gamma=
\displaystyle\frac d2 + \varepsilon$, $(\varepsilon\ll 1)$ we 
have for the index $p$ in (\ref{eg})
\begin{equation}
p_{\varepsilon}=\displaystyle\frac{2-d+2\varepsilon}{2}\,.
\label{pcrit}
\end{equation}
This  means that (a) when  $d=1$ the energy always decays with  
$p\simeq 1/2$, (b) when  $d=2$ the variance decays very slowly 
with $p=\varepsilon$,
(c) when  $d=3$ the variance increases with  $p\simeq -1/2$, and 
(d) when $d\gg 1$ then  $p\simeq (2-d)/2$.

This behavior of critical index  has a simple explanation if we consider 
the velocity field evolution in an elementary cell at the stage before
the interaction with other cells takes place (Section \ref{s:initial}). 
It follows from (\ref{Etas})
that the energy $E(t)$ of an elementary volume
varies as $E(t)\sim\psi_0^{\frac{d+2}{2}}t^{\frac{d-2}{2}}$.
When the power index $\gamma$ of potential probability  
distribution function 
is close to the critical value $\gamma_c$, then  probability
distribution function of the initial potential has 
a relatively slow decay and the
general evolution of the turbulence variance is determined by 
the cell with very high amplitude.
It is easy to see that at the critical point the time dependence of the
mean variance  (\ref{pcrit}) 
is the same as the time dependence of the  energy
of an initial cell (\ref{Etas}).

As $\gamma$ increases the interaction between the elementary cells
begins to play a  more and more important role. When $\gamma=d$ and  
$(d>2)$ the increase of variance $E(t)$ will switch to a 
decrease  with time.
When $\gamma\gg d$ the variance will decay as $t^{-1}$ which 
is close to the law
for the stretched exponential type of initial potential 
(see Section \ref{s:gumbel}).

It is easy to show from (\ref{py1}) that the probability  
distribution function of the coordinate of the absolute 
maximum of the potential  is also self-similar, i.e.
\begin{equation}
W({\bf y})=\displaystyle\frac{1}{L^d(t)}W_s\left(\frac{\bf y}{L(t)}
\right)\,,
\label{WYself}
\end{equation}
where $W_s(z)$ is the dimensionless form of probability distribution
function
\begin{equation}
W_s(z)=A \int\limits_0^{\infty}\displaystyle\frac{e^{-\Theta}d\Theta}
{\left[\displaystyle\frac{1}{\Theta^{\frac{2}{2\gamma-d}}}+z^2\right]
^{\gamma}}\,,
\label{WZG}
\end{equation}
$A$ is normalization  coefficient depending on $d$ and  $\gamma$.
In (\ref{WYself})   $L(t)$ is the external scale of turbulence and 
\begin{equation}
L(t)=\displaystyle\frac{(2 C_{d})^{\frac{1}{2}} 
(t H_*)^{\frac{\gamma}{2\gamma-d}}}{L_0^{\frac{d}{2\gamma-d}}}\sim
L_0\left(\frac{t}{t_{nl}}\right)^{\frac{\gamma}{2\gamma-d}}\,,
\label{LGT}
\end{equation} 

The  distribution (\ref{WZG}) decays  $\sim z^2$ at small distances and by
the power law at large distances:
$W(z)\simeq \displaystyle A/z^{2\gamma}$. From this we see again
that the  mean energy is finite when  
$\gamma > \gamma_{cr}=\displaystyle\frac d2 +1$, which is also  
the condition that the mean energy
in separate cells (\ref{Etas}) 
$E(t)\sim \int H^{\frac{d+2}{2}}f'(H)dH$ be finite.
Note that the mean initial energy 
(\ref{Et}) $E(0)=\langle E_0 \rangle /L_0^d
\sim \int H^2 f'(H)dH$ is finite if $\gamma >2$ for all value of $d$.

\subsection{Compactly supported potential: Weibull class }

\label{s:weibull}

Let us now assume that the initial potential is bounded by some value $H_m$
($\psi_0(y)\leq H_m$ in all cells) and that the cumulative probability
distribution function has a power law
behavior in the neighborhood of the maximum, i.e.
\begin{equation}
f(H)=\left(\displaystyle\frac{H_M-H}{H_*}\right)^{\alpha}
\,,\;\;H<H_M
\label{fHB}
\end{equation}
and is zero for $H>H_M$.
By the same procedure as above (see Section \ref{s:frechet}) 
it follows that the cumulative probability distribution function 
of potential has the form (\ref{qpp}) with
\begin{eqnarray}
N(H,t)=\displaystyle\frac{(H_M-H)^{\frac{2\alpha+d}{2}}(2t)^{d/2}}{L_0^d H_*^
{\alpha}}\cdot C_{d}\,,
\nonumber\\
C_{d}=\frac{2^{\frac{d+1}{2}}\pi^{d/2}}{\Gamma(d/2)}\int\limits
^1_0 r^{d-1}(1-r^2)^{\alpha}dr\,.
\label{NBsel}
\end{eqnarray}
and  is self-similar for  arbitrary dimension $d$ and power
index $\alpha$. The mean potential  tends to the maximum $H_M$  
according to the power law 
\begin{equation}
\langle  {\psi} \rangle = 
H_M - \tilde c H_* (t_{nl}/t)^{\frac{d}{2\alpha+d}}\,.
\label{HmeanB}
\end{equation}
Here $\tilde c$ is a positive numerical coefficient,
and $t_{nl}$  is the nonlinear time  
$t_{nl}$ (\ref{tnlr}), where 
the constant $C_d$ is determined by the expression (\ref{NBsel}).

From (\ref{HmeanB}) it follows that the energy always decreases 
according to the power law
\begin{equation}
E(t)=2\displaystyle\frac{\partial\langle {\psi} \rangle}{\partial t}\sim t^{-p}\,,
\;\; p=\frac{2(\alpha+d)}{2\alpha+d}\,.
\label{EB}
\end{equation}

It follows from (\ref{py1}),(\ref{fHB}),(\ref{NBsel}) that 
the p.d.f. of the   
absolute maximum coordinate ${\bf y}$ is self-similar
(\ref{WYself}) with the spatial scale $L(t)$
\begin{equation}
L(t)\sim L_0(t/t_{nl})^{\frac{\alpha}{2\alpha+d}}\,.
\label{LB}
\end{equation}
The general result for the length scale evolution for systems in the
Weibull class has actually been noted in \cite{NewmanS99} (although
not explicitly derived).
It is easy to see that the energy (\ref{eg}),(\ref{EB}) and 
the external  scale (\ref{LGT}),(\ref{LB}) of the turbulence 
are described by the same expressions for a  distribution 
having a power--law tail and for a  compactly supported distribution 
if we set $\alpha=-\gamma$ in  (\ref {fHB}) $(\gamma <0)$.

The form of the distribution of the absolute maximum coordinate 
is now defined by the integral
\begin{eqnarray}
W_s(z)=A\int\limits^{\infty}_{\theta_*(z)}\left( \theta^{\frac{2}
{2\alpha+d}}-z^2\right)^{\alpha} e^{-\theta}d\theta\,, 
\nonumber\\
\theta_*(z)=z^{2\alpha+d}\,.
\label{WZB}
\end{eqnarray}
It can be seen that this distribution decays exponentially 
at large $z$
\begin{equation}
W_s(z)\sim e^{-z^{2\alpha+d}} z^{\alpha(2-d-2\alpha)}\,.
\label{wzbas}
\end{equation}

It is interesting to compare the evolution of Burgers turbulence 
for a compactly 
supported initial potential with the evolution of turbulence for a Gaussian
initial potential having scaling properties with dimension $h$, such that 
$0\geq h > -1$ \cite{BeKr98}. For  values of $h$ 
that satisfy this condition the initial potential correlation
function has a singularity at the origin. In \cite{BeKr98} a special
class of random solution of the one-dimensional Burgers equation 
was constructed.
For this solution the self-similar behavior is true at all times 
and  not just  asymptotically. Such scaling behavior 
corresponds to the characteristic 
length $L(t) \sim t^{1/(2-h)}$. Comparing this result with (\ref{LB}) 
we see that for $\alpha=-1/h$ and $\alpha > 1$ the evolution of turbulence
with a compactly supported initial potential tends asymptotically to
the self-similar solution described in  \cite{BeKr98}.
 
From (\ref{EB}) we see that when probability distribution function 
of initial potential decays very rapidly ($\alpha\to\infty$) 
the mean energy decays as $t^{-1}$ just as it does  for the
stretched exponential type of initial potential.

The special case where $\alpha=1$ is equivalent to the uniform behavior
of the probability distribution function for the 
initial potential over  $H<H_M$, that is to say: 
$W(H)=-\displaystyle\frac{\partial f(H)}{\partial H}=H_*^{-1}$.
In this case the energy  $E(t)\sim t^{2(1+d)/(2+d)}$ which is
equivalent to the asymptotic law for the uniform distribution of
the initial potential \cite {Newman97}.

The limit $\alpha\to 0$ when the energy decays 
as $t^{-2}$ in all dimension $d$ is also interesting. In this case  
the probability distribution function of the
coordinate (and of the velocity) has the universal form
$W_s(z)\sim\exp{(-z^d)}$ for all $z$.
It easy to see that for $\alpha\to 0$ the integral scale of the turbulence
does not increase with time.
The case of  $\alpha\to 0$  has a simple explanation if we rewrite 
the cumulative probability distribution function for the 
initial potential corresponding  in the form
\begin{eqnarray}
f(H)=PE(H_M-H)
\,,
\nonumber\\
W(H)=-\displaystyle\frac{\partial f(H)}{\partial H}=
P\delta(H-H_M)\,.
\label{fdelta}
\end{eqnarray}
Here $E(z)$ is the unit function, $\delta(z)$ is the delta-function, and
$P$ is the probability that in some cells the  potential amplitude  is
exactly equal to $H_M$. This probability may by rewritten as
$P=(L_0/L_P)^d$ where $L_P$ is the typical distance between  cells when
the initial potential is equal to $H_*$ and  $L_0$ is the size of the cells.
It is easy to see that in this case the universal behavior of energy decay
$E(t)\sim t^{-2}$ is due to the fact that after some
intermediate stage we have ``frozen''  turbulence.
This  means that the spatial structure of the velocity will not 
change in time and only the amplitude of the velocity will decay like 
$\Delta L/t$, where $\Delta L$ is the random distance 
between two cells with
equal initial potential, i.e. ($H_i=H_j=H_M$).
In the case $d=1$   the velocity will be 
a sequence of the  lines $v=(x-x_i)/t$ with the immobile shocks at 
$(x_{i+1}-x_{i})/2$, where $x_i$ is the position of the point where
$\psi_i=H_M$. This case was considered in \cite{EsipovNewman,Esipov}
but with  some other assumptions and with other tools.
In arbitrary dimension $d$ the energy decays as $L_P^2/t^2$, and the
probability distribution function of the Lagrangian coordinate
and velocity  will have the stationary
form $W(z)=A e^{-z^d}$ with the scale $L\sim L_P\sim L_0 P^{-1/d}$.
For $d=2$ this distribution has a Gaussian form.
In conclusion we note  that this asymptotic ``frozen'' behavior for
bounded initial perturbation of potential takes place for arbitrary $\alpha$
when  $d\to\infty$.

\subsection{Stretched exponential tail \\
of the initial potential:  Gumbel class}

\label{s:gumbel}

Let us  now assume  that the cumulative probability 
distribution function of the initial potential (\ref{fmax}) 
has a stretched exponential tail
\begin{equation}
f(H)=(H/H_p)^{\alpha}\exp{[-(H/H_*)^{\beta}]}\,.
\label{PHE}
\end{equation}
For initial conditions of this  type of the distribution  potential 
is localized in a narrow region $\Delta H$ near the mean value
$H_0\gg H_*$.
In the integral (\ref{nint}) over ${\bf y}$  we can take into account 
only the quadratic term in the exponent and 
obtain the following expression
\begin{equation}
N(H,t)=\displaystyle\left(\frac{H}{H_p}\right)^{\alpha}\left[
\frac{2\pi t H_*^{\beta}}{\beta H^{\beta-1}L_0^2}\right]^{d/2}
e^{-(H/H_*)^{\beta}}\,.
\label{NHE}
\end{equation}

Let us introduce the dimensionless potential $z$ so that
\begin{equation}
H=H_* h_0 \left(1+\displaystyle\frac{z}{h_0^{\beta}\beta}\right)\,,
\label{HZ}
\end{equation}
where $h_0$ is a solution of the transcendental equation $N(H_*h_0)=1$:
\begin{eqnarray}
h_0^{\frac{2\alpha+d(1-\beta)}{2}}\left(\displaystyle\frac{t}{t_{nl}}\right)^
{d/2} e^{-h_0^{\beta}}=1\,,
\nonumber\\
h_0\simeq\left(\frac{d}{2} \ln \frac{t}{t_{nl}}\right)^{1/\beta}\,.
\label{TREQ}
\end{eqnarray}
Here $t_{nl}$ is the nonlinear time
\begin{equation}
t_{nl}=\displaystyle\frac{L_0^2}{2\pi H_*}\left(\frac{H_p}{H_*}\right)
^{2\alpha/d}\,.
\label{tnle}
\end{equation}
The variable $z$ in  (\ref{HZ}) has a universal double--exponential 
distribution at $t\gg t_{nl}$
\begin{equation}
F(z)=e^{-e^{-z}}\,.
\label{dexp}
\end{equation}
When $t\gg t_{nl}$ the distribution of the potential $\psi$ is concentrated
in the narrow region 
$\Delta H/H_0\sim h_0^{-\beta}\sim [d \ln t/t_{nl}]^{-1}$
near the mean value $\langle H\rangle =H_*h_0\sim (\ln t)^{1/\beta}$.
From (\ref{ephi}) we have the following expression for the energy of
turbulence
\begin{equation}
E(t)\simeq 2H_*\displaystyle\frac{\partial h_0}{\partial t}=
\frac{2H_*(d/2)^{1/\beta}}{t\beta}\left(\ln t/t_{nl}\right)^{\frac{1-\beta}
{\beta}}\,.
\label{EE}
\end{equation}
From (\ref{EE}) we see that the energy decays 
according to the universal law
$E(t)\sim t^{-1}$ with some logarithmic correction: the decay is 
faster if $\beta>1$
(tail of probability distribution function $f(H)$ decays faster then
the exponential law), and slower if  $\beta<1$.
It must be stressed also that the law of decay does not depend on the
dimension of the space and so is  the same as in one-dimension \cite{Kida}.

The probability distribution function of absolute maximum 
coordinate ${\bf y}$ and the
velocity  ${\bf v}$ (\ref{vxy}) are Gaussian 
and the variance of each component 
is given by
\begin{eqnarray}
\langle y_i^2 \rangle=L^2(t)=\displaystyle\frac{tH_*}{\beta h_0^{\beta-1}}
\sim\frac{t}{(\ln t/t_{nl})^{\frac{\beta-1}{\beta}}}\,,
\nonumber\\
\langle v_i^2 \rangle = L^2/t^2\,.
\label{EL}
\end{eqnarray}

The two-point probability distributions of the velocity and correlation
functions were found in \cite{GMS91,GurbatovSaichev1984}
for the special case ${\beta}=2$. The shape of the two-point 
probability distribution function of the longitudinal 
component is the same in the 
space of different dimensions, and coincides with the analogous p.d.f. of 
the one--dimensional Burgers turbulence. The transverse velocity 
components, unlike the
longitudinal ones,  are statistically independent in different cells 
and have a Gaussian probability distribution inside them. 
These results may be extrapolated
to the arbitrary stretched exponential tail of the initial potential.

\section{Concluding remarks}
\label{s:remarks}

The present work has considered the evolution of random solutions of
the unforced Burgers and  equations in d--dimensions in the limit 
of vanishing viscosity. The main statistical assumption is the independence
of the initial velocity potential $\psi_0 $ in different cells.

We show that the asymptotic behavior of the turbulence at large times 
is determined by the tail of cumulative initial potential 
probability distribution function
$F(H)=1-f(H)$. We show  that  three classes of initial distribution
lead to the  self--similar evolution of the turbulence at large times.
In the theory of extremes these limiting distributions are known 
as (a) Frechet class  when $f(H) \propto  H^{-\gamma}$; 
(b)  Weibull class  when 
$ f(H) \propto  (H_{max} -H)^{-\gamma} , \gamma\le 0$; 
and (c) Gumbel class  when  $f(H) \propto \ exp (-H^{\beta})$. 
One can find in \cite{LeadbetterLindgrenRootzen} more general 
conditions  which are necessary and sufficient 
for the probability 
distribution function $f(H)$ to belong to each of three 
types.

We show that  the mean potential (mean height of the surface) 
increases with time in cases (a),(b) according to the power law 
$\langle \psi \rangle \propto At^r , r=d/(2 \gamma-d)$ 
($A<0$ for $\gamma \le 0$) while in case (c) it increases   
logarithmically according to 
$\langle \psi \rangle \propto (\ln t)^{1/{\beta}}$.
For the Gumbel class  the distribution of 
potential is localized in a narrow region near the mean value.
For the Weibull class a p.d.f. of potential exists for 
$\gamma > d/2$ and the  mean value of $\psi$ is 
finite for $\gamma > d/2+1$.
The mean square gradient of the surface (turbulence energy) 
$E(t)=\langle (\nabla {\psi}({\bf x},t))^2\rangle=
\langle {\bf v}^2({\bf x},t)\rangle$ has the power-law  dependence  
$ E(t) \propto t^{-p}, p=2(\gamma-d)/(2\gamma-d)$ in cases (a) and (b) 
and decays according to 
$ t^{-1} (\ln t)^{(1-\beta)/{\beta}}$ in case (c).
For relatively slow decay of the initial probability distribution of potential 
$1+d/2 <\gamma <d$ and dimension of the space $d>2$ the energy $E(t)$ 
increases with time. 

We show that the p.d.f. of the velocity is 
self--similar with the scale $L(t)/t$, where $L(t)$ 
is the external scale of the turbulence.
In  cases (a) and (b) the external scale of turbulence increases 
according to the pure power law
$L(t) \propto t^{m}, m=\gamma /(2\gamma-d)$ while in  case (c) 
it increases like
$L(t) \propto t^{1/2} (\ln t)^{(\beta-1)/{\beta}}$. One can see 
that for  fast decaying initial distributions of potential 
($|{\gamma}|\to\infty$) in classes (a) and (b)) the law of 
external scale increase and the law of energy decay 
tends to the corresponding laws of the  Gumbel class. 
In the special case $\gamma \rightarrow{-0}$ (class b)
we have the ``frozen'' turbulence which means that the structures of the 
potential and velocity fields conserve $L(t)=const$  
and the amplitude of the velocity increases as $t^{-1}$. 
We note that for the compactly supported potential 
this ``frozen'' behavior takes 
place for arbitrary $\gamma$ when  $d\to\infty$.

We now discuss what influence  finite viscosity has on the asymptotic 
behavior of the Burgers turbulence at large times.  For  large 
initial Reynolds number ($Re_0\gg 1$) we still  have the 
cellular structure of the turbulence at relatively large times. For  
finite Reynolds number  it will be characterized by two scales, namely 
the external scale $L(t)$  and inner scale 
$\delta$ the latter being a  typical width of the shock surface 
\cite{GMS91}. Owing to  viscosity the inner scale 
increases as $\delta \sim \nu t/L$. However due to the increase of 
$L(t)$ the relative width of the shock 
$\delta (t)/L(t) \sim  \nu t/L^2 (t) \sim Re^{-1}(t)$  may either 
decrease or increase with time. Here we write the  Reynolds 
number as  $Re(t)=V(t)L(t)/{\nu}=L^2 (t)/\nu t$, 
since the local slope of the velocity is $1/t$ and the 
maximum velocity is of order $L(t)/t$. 

It is easy to see that for  
classes (a) and (b) we have a power law for the Reynolds number,
namely $Re(t) \propto t^r , r=d/(2 \gamma-d)$. 
Thus for the Frechet class 
($\gamma>d/2$) the Reynolds number increases with time and hence even 
when the viscosity is  finite we have  the strong 
nonlinear stage of evolution at large times. 
For the Weibull class ($\gamma \le 0$) 
the Reynolds number decreases with time and at large times the evolution of 
the turbulence will be determined only by the linear diffusion. 
This is shown in  \cite{Newman97}
for the case of a uniform distribution of the initial potential. 
In \cite{Newman97}  it was stressed that although one finds a universal
power law growth for the energy decay (for a system with a bounded
and flat initial distribution), the velocity-velocity correlation
function has a more complicated form, and there is {\it no simple dynamical
scaling in the system} in the `non-linear regime'.
For the Gumbel class   
($f(H) \propto \exp(-H^{\beta})$) the Reynolds number 
$Re(t) \propto (\ln t)^{(1-\beta)/{\beta}}$. Thus we may expect that for
$\beta <1$ we have  conservation of the cellular structure at large times. 
For $\beta >1$ the nonlinear evolution is only an intermediate 
asymptotic which changes at large times to linear decay. 
This effect is considered   in 
\cite{GMS91,AMS94} in the case of a Gaussian initial 
perturbation.

Let us now move to  the case when the initial potential is long--range
correlated or non--homogeneous. We  consider first  one-dimensional 
turbulence, assuming that the initial velocity is homogeneous
with a spectrum  $E_v (k) \sim \alpha ^2 |k|^n$ at small
wavenumbers $k$ and falling off quickly at large wavenumbers.
For the spectrum of potential we have 
$E_{\psi} (k) \sim \alpha ^2 |k|^{n-2}$. For a Gaussian velocity
it is shown in \cite{GSAFT97} that there are three regions of $n$ 
with different behavior of turbulence. When $-1<n<1$  
the long-time evolution of the spectrum is self-similar and the 
external scale of turbulence 
increases as $L(t) \propto(\alpha  t)^m , m=2/(3+n)$ and is determined
by the ``amplitude'' of the  large scale component ${\alpha}$. 
When $1<n<2$ the spectrum  has three scaling regions\,:
first, a $|k|^n$ region at very small $k$ with a time-independent
constant, associated with  long-correlated regions in physical space,
second, a $k^2$ region at intermediate wavenumbers which is 
related to the self-similarly evolving ``inner
region'' in physical space and, finally, the usual $k^{-2}$ region,
associated to the shocks.  The growth of the  external scale is now 
determined by two integrals of the initial spectrum  and 
$L(t) \propto t^{1/2} (\ln t)^{-1/4}$.
Switching wavenumber from the $|k|^n$ to the $k^2$
region tends to zero faster than the energy wavenumber
$\sim 1/L(t)$ and asymptotically we have the self--similar 
evolution of the spectrum. For $n>2$,
long-time evolution is also self-similar and 
$L(t) \propto t^{1/2} (\ln t)^{-1/4}$. Thus for a Gaussian 
perturbation we have one critical index $n$ for the behavior of 
external scale and energy. When $n=1$ the index of the power law 
dependence of $L(t)$ continuously transforms from $m=2/(3+n)$, ($n<1$) 
to the $m=1/2$  index of the leading term in the region $n>1$.

Let us assume that for the multi--dimensional Burgers turbulence the 
initial potential is isotropic and has a power--law dependence at small
wavenumbers, i.e.  $E_{\psi} (k) \sim \alpha ^2 |k|^{n-2}$. The variance
of the potential is determined by 
$\langle {\psi}^2 \rangle \sim \int\limits_0^{\infty}E_{\psi} (k) k^{d-1}dk$  
and is finite when $n>2-d$.
For $2-d <n<-d$ its structure function is
$d_{\psi}({\bf x})=\langle \left(\psi_0({\bf x})-\psi_0(0)\right)^2\rangle 
\sim |{\bf x}|^{2-n-d}$. Using the rescaling of the structure function in
(\ref{psimax}) we see that the external scale increases as
$L(t) \sim (\alpha t)^m, m =2/(2+n+d)$ 
\cite{GMS91,VDFN94}. This
law does not depend on the p.d.f. of the initial potential.
Introducing the Reynolds number directly through the Hopf-Cole solution
\cite{GMS91} 
$Re(t)=L^2 (t)/{\nu=V(t)L(t)/{\nu}} t$ we see that in this region of
$n$ the Reynolds number increases with time 
according to $\propto t^{(2-n-d)/(2+n+d)}$.
It  means that at late stage  the turbulence has a  
strong nonlinear cellular structure.
 
Considering the case of independent amplitudes in different cells, 
we see that the correlation function of the initial potentials 
equal to zero if the distance between the points is greater then 
the size of the cell.
Thus, the discrete model of the initial conditions  
considered in the present paper
is equal to the  uncorrelated potential with $n \ge 2$.  
Here we show that the laws of  evolution
of the external scale $L(t)$ (\ref{LGT}),(\ref{LB}),(\ref{EL}) 
and the energy 
$E(t)$ (\ref{eg}),(\ref{EB}),(\ref{EE})   are very sensitive 
to the  tail of the  p.d.f. of the initial potential. 
For the stretched exponential
tail (i.e. the Gumbel class) the leading term for the evolution of the  
external scale is $t^{1/2}$.
Thus we can expect that in  multi-dimensional
turbulence the long--range correlation of the potential 
for $2-d<n<2$ does not
play an important role for the evolution of the external scale and energy 
and we have only one critical index $n$. This is shown in
\cite{MSW95,GSAFT97}
for the one dimensional  case with a Gaussian potential.
When $n=2-d$ the universal index of the power law 
dependence of $L(t)$ continuously transforms from $m=2/(2+n+d)$, ($n<2-d$) 
to the index  $m=1/2$ for the  leading term  in the region where $n>2-d$. 
In this case we have also the same critical index
$n=2-d$   for the energy. In the interval $2-d<n<2$
we have  conservation  of the  velocity spectrum $E_v (k) \propto |k|^n$ at
very small wavenumbers. But this small region is not significant for the
energy of the turbulence and asymptotically the spectrum of the velocity
tends to the self--similar evolution.

We have a much more nontrivial situation when the potential distribution 
has a with power--law tail (Frechet class) or  when it is a 
compactly supported potential (Weibull class). 
For $n<2-d$ we find, using  the scaling
properties of solution (\ref{psimax}), that the external scale increases as
$L(t) \sim (\alpha t)^m, m =2/(2+n+d)$.  
Thus the evolution of $L(t)$ and the energy $E(t) \sim L^{2}(t)/t^2$ do not 
depend on the p.d.f. of the initial potential. For the uncorrelated potential
$n \ge 2$  the power indexes laws  of the scale evolution
$m={\gamma}/(2{\gamma}-d)$ and energy evolution 
$p=2({\gamma}-d)/(2{\gamma}-d)$
depend on  $\gamma$. The case of a homogeneous continuous potential
with a power index  $2-d<n<2$ is equal to the existence of 
long--range correlation of the the potential amplitudes
in  cells. We can expect that in the region $2-d<n<2$
the long--range correlation of potential influences the energy decay.

\vspace{2mm} \par\noindent {\bf Acknowledgments.}
We have benefited from discussion with  
U.Frisch, A.Saichev, A.Noullez and J.Tully.
This work was supported by 
INTAS through grant No 97--11134, by
RFBR through grant 99-02-18354. S.Gurbatov thanks the
French Ministry of Education, Research and Technology for support 
during his visit to the  Observatoire de la C\^ote d'Azur.

\end{document}